\documentclass[aps,prl,twocolumn,groupedaddress,amsmath,amssymb]{revtex4-1}
\usepackage{graphicx}  % needed for figures
\usepackage{dcolumn}   % needed for some tables
\usepackage{bm}        % for math
\usepackage{verbatim}   % for math
\usepackage{hyperref}
\usepackage{color,soul}

\begin{document}

\title{Curvature-controlled defect dynamics in active systems }
\author{Sebastian Ehrig}
\email{Electronic address: sebastian.ehrig@mpikg.mpg.de}
\affiliation{
   Department of Biomaterials,
   Max Planck Institute of Colloids and Interfaces,
   14482 Potsdam,  Germany
 }
\author{Jonathan Ferracci}
\affiliation{
   Department of Biomaterials,
   Max Planck Institute of Colloids and Interfaces,
   14482 Potsdam,  Germany
   }
   \author{Richard Weinkamer}
\affiliation{
   Department of Biomaterials,
   Max Planck Institute of Colloids and Interfaces,
   14482 Potsdam,  Germany
   }
	\author{John W. C. Dunlop}
	\email{Electronic address: john.dunlop@mpikg.mpg.de}
\affiliation{
   Department of Biomaterials,
   Max Planck Institute of Colloids and Interfaces,
   14482 Potsdam,  Germany
   }
	
\date{\today}

\begin{abstract}
We have studied the collective motion of polar active particles confined to ellipsoidal surfaces. The geometric constraints lead to the formation of vortices that encircle surface points of constant curvature (umbilics). We have found that collective motion patterns are particularly rich on ellipsoids with four umbilics where vortices tend to be located near pairs of umbilical points to minimize their interaction energy. Our results provide a new perspective on the migration of living cells, which most likely use the information provided from the curved substrate geometry to guide their collective motion. 
 \end{abstract}

\maketitle

%\section{Introduction}
\textit{Introduction} - Active particles are known to spontaneously form complex dynamic patterns at length scales ranging from the molecular \cite{decamp2015orientational}, to the cellular \cite{szabo2006phase, trepat2009physical} up to macroscopic patterns seen in flocking birds \cite{cavagna2013diffusion}, schooling fish \cite{calovi2014swarming} or humans in crowded environments \cite{silverberg2013collective, karamouzas2014universal}. The key feature of these active systems is the constant energy input on each individual unit, which renders the system completely out of equilibrium. Collective phenomena in such active systems have been successfully described using so-called self-propelled particle models \cite{vicsek1995novel} that are limited to close neighbour interactions only \cite{vicsek2012collective}. In unconstrained 2D and 3D systems these models display self-organised pattern formation resembling experimental observations \cite{vicsek2012collective}. The behaviour of active particles confined to a surface has been mainly studied on planar surfaces of zero Gaussian curvature. It is known however, that the presence of intrinsic surface curvature frustrates local order giving rise to novel physics \cite{irvine2010pleats}, as has been shown for 2D fluids confined to curved surfaces \cite{reuther2015interplay}. As a consequence of the Poincar\'{e}-Hopf theorem, for instance, it is not possible to have continuous fluid flow on the entire surface of a sphere, which requires the presence of two +1 defects (vortices) \cite{kamien2002geometry}. The effect of non-zero Gaussian curvature on self-propelled particles remains poorly understood, with only a few recent examples studying the effect of spherical constraints \cite{sknepnek2015active, khoromskaia2016vortex}. In living systems, cells are influenced by surface curvature as demonstrated by cell movements in the developing corneal epithelium leading to vortex patterns \cite{collinson2002clonal} or by the coordinated collective migration of cells during embryonic development \cite{keller2008reconstruction}. The emergent behaviour of moving cells is not only the result of intercellular interactions, but is crucially influenced by geometrical constraints \cite{szabo2006phase, wan2011micropatterned, duclos2014perfect}. The aim of the current work is to investigate the impact of non-constant Gaussian curvature constraints on the collective behaviour of self-propelled particles. Our restriction of the geometry of the surfaces to ellipsoids allows an analysis of how geometrical cues (represented by the umbilical points of the surface, Fig. 1c) effectively interact with defects in the director-field (e.g., vortices). The strong coupling between vortex position and umbilical points demonstrates the importance of surface geometry on the emergence of patterns in active systems. This work could have significant implications in understanding collective phenomena especially in the context of growing tissues, where cell movements are constrained to constantly changing surfaces.

%\section{Methods}
\textit{Methods} - We use a Vicsek type model \cite{vicsek1995novel} of \textit{N} spherical active particles of radius $\sigma$ confined to the surface of an ellipsoid with principle axes x, y, z. Particles are self-propelled (moving with a scalar self-propulsion term $v_{0}$) and are polarized (being oriented towards the direction \textbf{\textit{n}}). Particle interactions occur via a short ranged linear force potential consisting of short ranged repulsive forces $\textbf{\textit{F}}_{rep}$ and attractive forces $\textbf{\textit{F}}_{adh}$ from neighboring particles scaled by the mobility parameter $\mu$. The overdamped equations of motion for particle \textit{i} are described by:

\begin{equation}
\frac{d\textbf{r}_{i}(t)}{dt}=v_{0}\textbf{n}_{i}(t)+\mu\sum\limits_{j=1}^N\textbf{F}(\textbf{r}_{i},\textbf{r}_{j})
\end{equation}

where $r_{i}$ is the position of particle \textit{i} and $\textbf{F}(\textbf{r}_{i},\textbf{r}_{j})$ is the short ranged linear force potential (Fig. 1b) given by \cite{szabo2006phase}

\begin{equation}
		\textbf{F}(\textbf{r}_{i},\textbf{r}_{j}) =\textbf{e}_{i,j}
  \begin{cases}
    F_{rep}\frac{d_{ij} - R_{eq}}{R_{eq}}, & \text{if $d_{ij} < R_{eq}$ }\\
    F_{adh}\frac{d_{ij} - R_{eq}}{R_{0} - R_{eq}}, & \text{if $R_{eq} \leq d_{ij} \leq R_{0}$ }\\
		0, & \text{if $R_{0} < d_{ij}$ }
  \end{cases}
\end{equation}

where $\textbf{e}_{i,j}=(\textbf{r}_{j}-\textbf{r}_{i})/|\textbf{r}_{j}-\textbf{r}_{i} | ,d_{i,j}=|\textbf{r}_{i}-\textbf{r}_{j} |$, $F_{rep}$ and $F_{adh}$ are the values of the maximum repulsive and attractive forces at $d_{ij}=0$ and $d_{i,j}=R_{0}$ respectively. In the presence of neighboring particles, the particle direction \textbf{\textit{n}} and direction of motion $\dot{\textbf{\textit{r}}}$ usually deviate and the particle direction \textbf{\textit{n}} realigns with the velocity $\dot{\textbf{\textit{r}}}$ according to:

\begin{equation}
\frac{d\textbf{n}_{i}(t)}{dt}=-\frac{\textbf{r}_{i}\times\dot{\textbf{r}_{i}}}{\tau\lVert\dot{\textbf{r}_{i}}\rVert}\times\textbf{n}_{i}+\xi
\end{equation}

where $\tau$ is the relaxation time and $\xi$ is angular noise described by a delta correlated Gaussian white noise term with zero mean, $\left\langle\xi(t)\xi(t')\right\rangle=\eta\delta(t,t')$.\\
Particle motion on the curved surface is performed by an unconstrained motion in the tangential plane followed by a projection onto the surface. In order to preserve the absolute values of the individual velocities, the particle velocity vector and orientation vector are rotated with respect to the angular difference between the surface normal of the initial and final tangent plane (Fig. 1a). Arbitrary surfaces are approximated with triangulated meshes generated via a custom mesh relaxation algorithm in \textit{Rhino}/\textit{Grasshopper} \cite{McNeel, Grasshopper}.

%\twocolumngrid
\begin{figure}
\includegraphics[width=\linewidth]{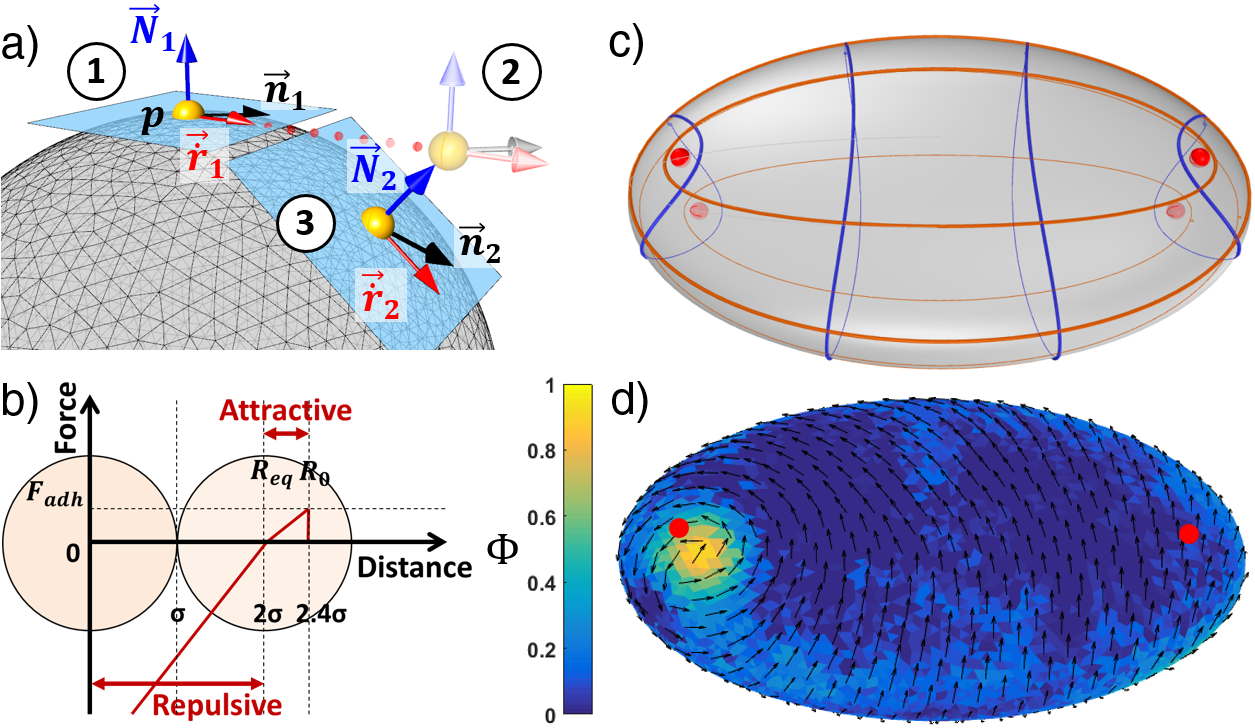}
\caption{\label{Geometry_and_definitions} Active particles confined to the surface of an ellipsoid. a) Particle motion on a triangulated surface is performed in two steps: unconstrained motion in tangential plane (1, 2) followed by a projection onto the surface (3). $\vec{N}_{1}, \vec{n}_{1}, \dot\vec{r}_{1}$ are the surface normal-vector, particle orientation-vector, and velocity-vector at point \textit{p} before the projection. b) Particles interact via a finite short-ranged repulsive/attractive linear force potential within a cut-off distance $R_{0} = 2.4\sigma$. c) Example of an ellipsoid with principal axis $x=4$; $y=2.5$; $z=1$ resulting in aspect ratios $x/z=4$ and $x/y=1.6$; lines of maximum and minimum principle curvatures are shown in blue and orange respectively. Points of constant normal curvature (umbilics) are highlighted as red spheres. d) Director-field (black arrows) and vortex order parameter ($\Phi$; color-coded) on the ellipsoidal surface shown in c) after 15800 time steps; red dots indicate positions of umbilics.}
\end{figure}
          
To test the influence of varying Gaussian curvature on pattern formation of self-propelled particles, we have performed particle simulations on two classes of ellipsoidal surfaces: (i) spheroidal and (ii) non-spheroidal. General ellipsoidal surfaces (shown in Fig. 1c) are characterised by their three principal axis x, y, z and have non-constant Gaussian curvature. Spheroidal ellipsoids are either prolate $(x/z=y/z<1)$ or oblate $(x/z=y/z>1)$. However, there are points on the surface which are “sphere-like”, i.e., where any direction is a principal direction, which are called umbilical points or umbilics. In contrast to the surface of a sphere, where every point is an umbilic, ellipsoidal surfaces have a finite number of umbilical points: having either 2 (spheroids) or 4 (non-spheroidal ellipsoids) (Fig. 1c). Simulations were performed on ellipsoids of varying aspect ratios (see Fig. 3) and all surfaces were scaled such that the surface area is always the same. Units of length, time and mass are defined in the model by specifying $R_{0}=1$, the relaxation time $\tau =1$, and the mobility parameter $\mu=1$. The model included $N=828$ particles at a fixed particle radius $\sigma=5/12$ and packing fraction $\varphi=1$ (defined as the ratio of the cross-sectional area of the particles to the total surface area of a reference sphere with radius $R_{SP}=6$, $\varphi=N\pi\sigma^{2}/4\pi\textit{R}_{SP}^{2}$). The interaction parameters between the particles were based on those used in [3]: $\textit{F}_{rep}=10$, $\textit{F}_{adh}=0.75$, $R_{eq}=5/6$ and $\eta=2(10^{-3})$. Values of the self-propelled velocities range from $v_{0}=0.1$ to $v_{0}=0.5$ and are chosen such that $v_{0}\ll\mu\textit{F}_{rep}$, therefore, the study is in the regime of low noise and low velocity and particles interact virtually as hard spheres. The mesh size was chosen to be inversely proportional to the local Gaussian curvature and much smaller than the particle radius resulting in typical numbers of surface triangles of 10 times the particle number. Particles are initially randomly distributed on the surface with random overlaps and random orientations. All simulations have been performed in \textit{Matlab R2015b} by solving the overdamped differential equations of motion (1) and (3) using a forward Euler integration method with a fixed time step of  $\Delta t=0.01\tau$ for a total of $2.5(10^{4}\tau)$ time steps.\\

%\onecolumngrid
\begin{figure}[h]
\includegraphics[width=\linewidth]{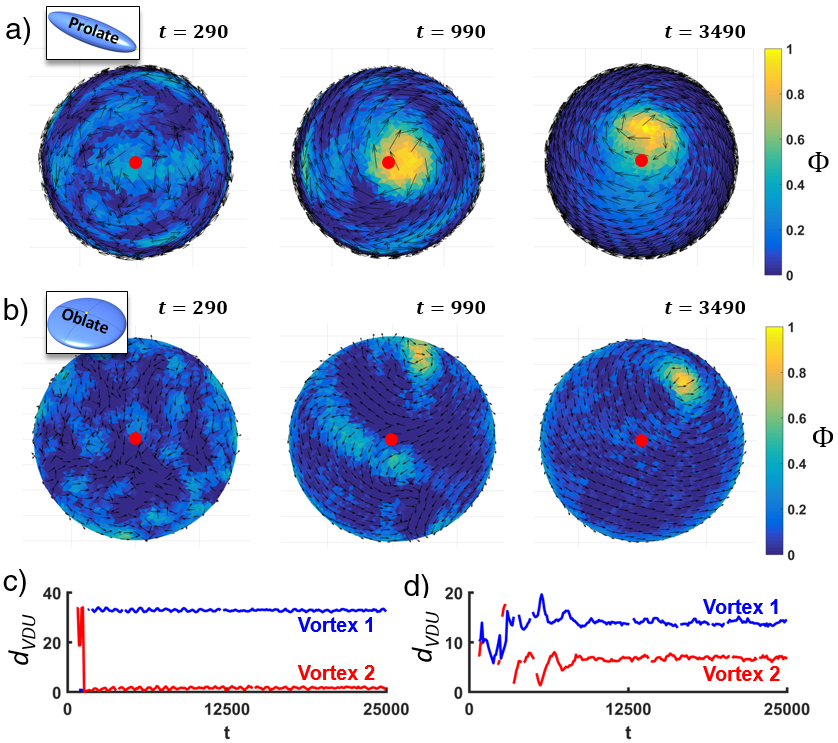}
\caption{\label{Evolution_of_the_VOP} Evolution of the vortex order parameter $\Phi$ on a prolate (a) and oblate spheroid (b) with aspect ratios $x/z=0.25$ and $x/z=4$ respectively. Vortices are quickly formed near umbilical points (red dots) where they maintain a constant geodesic distance ($d_{VDU}$, vortex distance to umbilic) between vortex center (color-coded) and umbilics (c, d) with a significantly smaller separation distance on prolate spheroids compared to oblate spheroids. The distance $d_{VDU}$ from vortex position (center of mass of $\Phi > 0.7$) to umbilic stabilizes almost instantaneously on the prolate spheroid (c) whereas it takes considerably longer on the oblate spheroid (d).}
\end{figure}
%\twocolumngrid

The directed motion of active particles and the spherical topology of the ellipsoid usually lead to the formation of two vortices (Fig. 1d, second vortex at the back of the ellipsoid). The position of the vortices on the surface was determined by adapting the 2D vortex order parameter (VOP) recently introduced by \cite{lushi2014fluid}, defined by 

\begin{equation}
\Phi=\frac{1}{1-2/\pi}\left(\sum\limits_{i}\left|\textbf{n}_{i}\cdot\textbf{t}_{i}\right|/\sum\limits_{j}\left\|\textbf{n}_{j}\right\|-\frac{2}{\pi}\right)
\end{equation}

where $\textbf{\textit{n}}_{i}$ is the orientation of particle \textit{i}, $\textbf{t}_{i}$ is the azimuthal unit vector to the tangent plane; $\Phi=1$ for purely azimuthal and $\Phi=0$ for pure radial orientations. The VOP has been evaluated at each vertex point of the triangulated surface including the first 3 shells of particle neighbours. The position of the vortex was then defined as the local center of mass of the calculated VOPs, for values above 0.7. We then evaluated the geodesic distance between the two vortices and between each vortex and each umbilical point ($d_{VDU}$, vortex distance to umbilical point), whereby $d_{VDU}$ is the distance between the vortex-center and the umbilical point. Geodesic distances are measured on the triangulated surfaces using the \textit{Toolbox Fast Marching} \cite{Peyre} which is an implementation of the Fast Marching algorithm introduced by \cite{sethian1996fast}.

%\section{Results}

\textit{Results} - In order to investigate the influence of the umbilical points on the dynamics of these defects, we performed simulations on (i) spheroidal and (ii) non-spheroidal ellipsoids. On spheroids the system of active particles showed a two-phase dynamic behaviour: on short time scales ($t < 1000$) two vortices form at opposite sides of the spheroid. This is followed by a transition period (Fig. 2c, d), in which these two vortices rotate around the surface normal at the umbilical points forming a stable motion pattern (supplemental material movies 1 and 2 \cite{Supplemental_Movies}). The snapshots of Figure 2 show the formation of a vortex (yellow region) close to an umbilical point (red dots) on prolate (Fig. 2a) - and oblate-spheroids (Fig. 2b) at three consecutive time-points. After their formation, vortices maintain an almost constant $d_{VDU}$ with a significantly smaller separation distance on prolate (Fig. 2c, suppl. mat. movie 1 \cite{Supplemental_Movies}) compared to oblate spheroids (Fig. 2d, suppl. mat. movie 2 \cite{Supplemental_Movies}). By systematically changing the aspect ratio of the spheroid (Fig. 3a), we found that for prolate spheroids the $d_{VDU}$ is smallest for large aspect ratios and decreases as spheroids become more elongated (low x/z). The same trend with aspect ratio can be observed for oblate spheroids however with higher $d_{VDU}$ values when compared to prolate spheroids of similar aspect ratios (a profile curve for $v_{0}=0.5$ is presented in Fig. 3c).

%\onecolumngrid
\begin{figure}[h!]
\includegraphics[width=\linewidth]{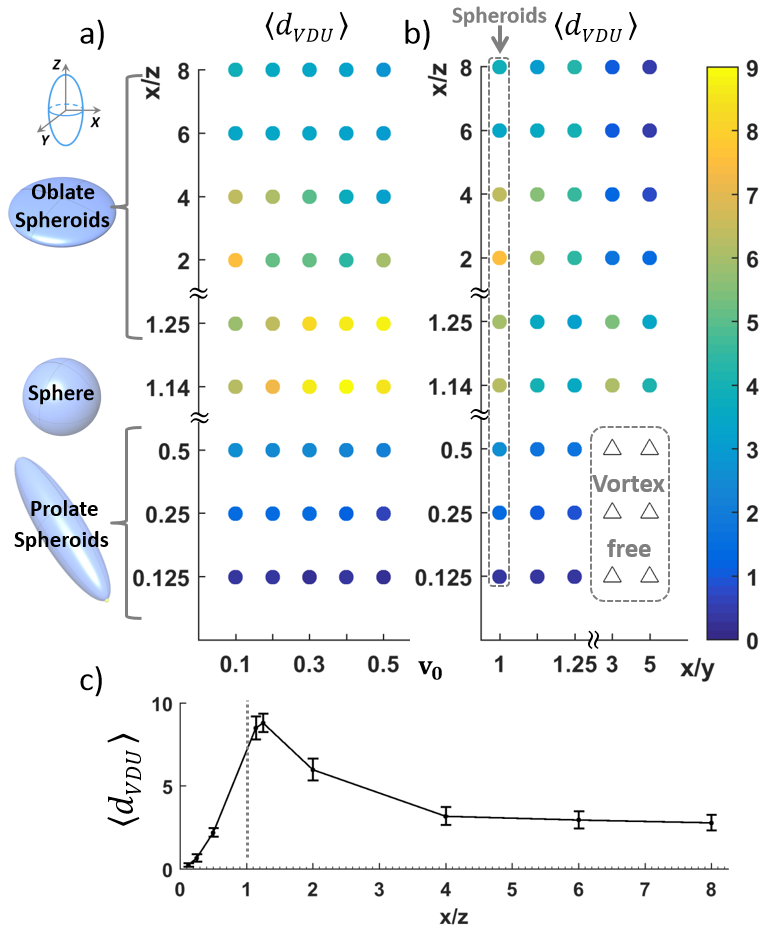}
\caption{\label{Evolution_of_the_VOP1} Mean distance of vortex center to umbilical points $\langle d_{VDU}\rangle$ for a) spheroids of different aspect ratios and velocities, and b) triaxial-ellipsoids of different aspect ratios at a constant particle velocity $v_{0}=0.1$. The $\langle d_{VDU}\rangle$ shown in a) increases with the aspect ratio of the spheroids. $\langle d_{VDU}\rangle$ obtains significantly smaller values (even for small aspect ratios) on prolate spheroids compared to oblate spheroids. A profile curve for $v_{0}=0.5$ is shown in c). On triaxial ellipsoids the $\langle d_{VDU}\rangle$ is also correlated with the aspect ratio with zones of stabilized-vortices and regions that are vortex free (depicted as triangles). Each data point was averaged over 10 independent simulations.}
\end{figure}

Particles interact via distant dependent forces that translate into an effective potential energy. In order to minimize this energy particles tend to move parallel to their neighbours. Hence, the total potential energy in the system will eventually transition to a lower energy state for long enough simulation times. As a result, particles distant from the poles of the spheroids and their umbilical points perform a collective motion which can be best described as band formation. In order to maximise the alignment of their velocities (to minimise the interaction energy) they move along geodesic paths and hence coherently move in one direction (suppl. mat. movies 1 and 2 \cite{Supplemental_Movies}). Depending on contingencies in the initial conditions of the simulation, the particle band structure can split into several sub-bands with opposite (i.e. counter-rotating) movement directions. These sub-bands were found to be stable over the length of the simulation (suppl. mat. movie 3 \cite{Supplemental_Movies}). The particle bands are a consequence of the spherical topology and are different from the high-density particle bands that occur in the Vicsek model for large system sizes. Here, high-density particle bands propagate perpendicular to their elongation direction parallel to the mean particle polarization and are surrounded by a background of mainly uncorrelated particles of low density. Such moving localized structures have been observed in two-dimensional euclidean \cite{gregoire2004onset, bertin2006boltzmann, bertin2009hydrodynamic, mishra2010fluctuations, ihle2013} and open three-dimensional space \cite{chate2008modeling}.

\begin{figure}[h!]
\includegraphics[width=0.9\linewidth]{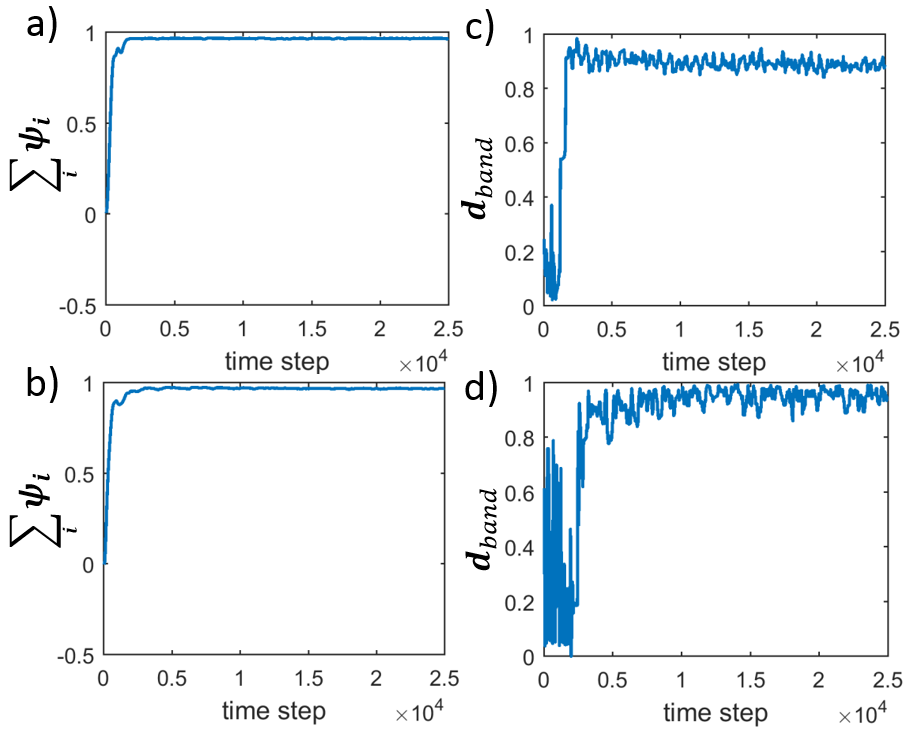}
\caption{\label{particle band structure} Measurements of the particle band structure for prolate (a, c) and oblate (b, d) spheroids of aspect ratio $x/z=0.25$ and $x/z=4$, respectively. The directed collective motion leads to a band formation that can be measured by summation of the local average of the dot product of each particle with its nearest neighbours ($\Psi_{i}$, local mean vector dot product of particle $i$)(a, b). The particle band quickly forms and was found to be stable over the entire simulation time (c, d). Particle band thickness ($\textbf{d}_{band}$) is defined as the distance between the boundaries of the regions that have $\Psi$ values below 0.8, normalized by maximum geodesic distance between the poles of the spheroids.}
\end{figure}

The particle bands on the ellipsoids are quickly formed as can be seen in the evolution of the total particle alignment (Fig. 4) and are stable over the entire simulation time. In order to explore the stability of these particle bands, we have tested the influence of system size for spheroids of two different aspect ratios ($x/z=0.25$ and $x/z=4$) at a constant particle density (Fig. 5). Such a system size scaling keeps the overall shape constant but changes the local Gaussian curvature. For particle simulations of up to $1.3(10^{4})$ particles we observe the same trend of vortex umbilical point distance as seen for small particle numbers. However, the time required to reach stability increases with the particle number and takes significantly longer on oblate spheroids (with pronounced fluctuating $d_{VDU}$ values) compared to prolate spheroids (Fig. 5). Furthermore, the $d_{VDU}$ seems to decrease with system size which might be related to the finite long range ordering of particles as well as the decrease in local Gaussian curvature. This would hence lead to a reduction in the vortex-vortex and vortex-umbilical point interactions. The high-particle density structures that occur in the Vicsek model were not observed for these larger surfaces, although we have only performed simulations with small particle numbers and low particle velocities. The main focus of the paper is to investigate the role of Gaussian curvature on active particles constrained to ellipsoids by systematically varying the aspect ratio. Further work will be required to understand the role of system size and hence the magnitude of curvature on pattern formation in active particle simulations (Fig. 5).\\
Two new dynamical features are observed in the collective motion on non-spheroidal ellipsoids. The first new feature is caused by the presence of four umbilical points, which causes a dynamic exchange of the two vortices between pairs of umbilical points that have a large geodesic distance. For low velocities ($v_{0}=0.1$) vortices encircle pairs of umbilical points resulting in oscillating values of the $d_{VDU}$ for both vortices (Fig. 6a, b, d, e, suppl. mat. movie 4 \cite{Supplemental_Movies}). Here, each vortex has the largest separation distance from the other vortex when both are in the vicinity of umbilical points (Fig. 6c). At higher velocities ($v_{0}=0.5$) the vortices become confined to regions of high Gaussian curvature between umbilics and the direction of the bulk particle motion becomes aligned with principle curvature directions (suppl. mat. movie 5 \cite{Supplemental_Movies}). The pairs of umbilical points that a vortex encircles can be exchanged during a simulation, however this exchange is coupled to the motion of the other vortex, as both vortices tend to maximise their separation distance.\\
The evolution of the vortex distance to each of the four umbilical points is correlated with the aspect ratio of the ellipsoid (Fig. 7). On prolate-like (black curves in Fig. 7) and oblate-like (blue curves in Fig. 7) ellipsoids with large aspect ratios and pairs of close umbilics, the $d_{VDU}$ quickly stabilizes. On ellipsoids with umbilics that are further apart (red curves in Fig. 7a, b), however, the $d_{VDU}$ exhibits stable oscillations after a longer transition phase. In each of these cases, two vortices form that maximize their separation distance (Fig. 7c) and are stable over the entire simulation time (Fig. 7d).\\
The second new feature offered by non-spheroidal ellipsoids was detected for flat prolate-like ellipsoids ($x/y>3$; $x/z<1$). In this case no stable vortices are formed (triangles in Fig. 3b). The formation of band-like collective motion is suppressed on these surfaces due to the highly curved edge which inhibits particle motion between the upper and lower surfaces of the “flattened” ellipsoids, thus constraining particle motion to either the upper or the lower surface (suppl. mat. movie 6 \cite{Supplemental_Movies}). In addition, for extremely flat prolate-like ellipsoids ($x/y>5$; $x/z<1$), the particles perform a collective oscillatory movement between the poles of the surface.

\begin{figure*}
\includegraphics[width=0.9\linewidth]{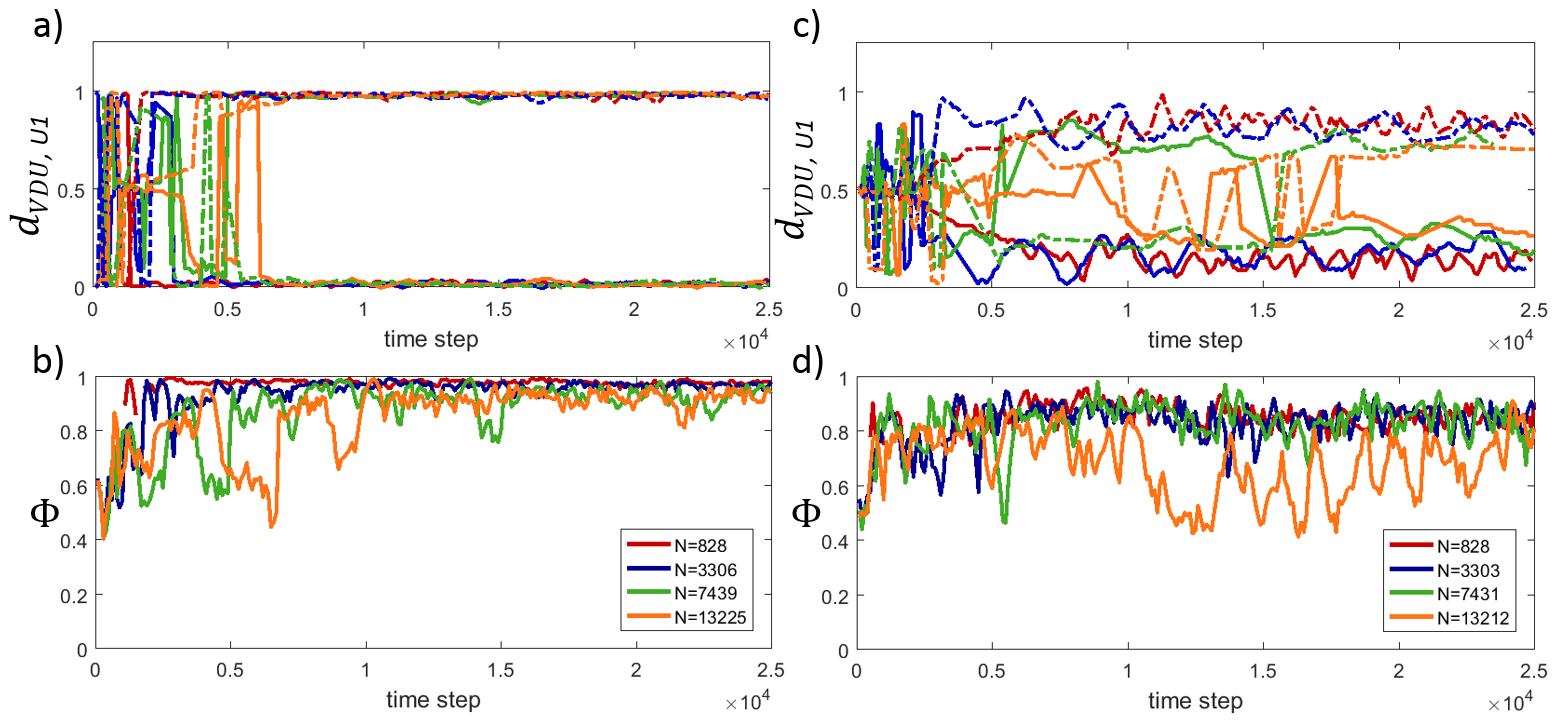}
\caption{\label{Evolution of VDU as a function of system size} Evolution of the vortex distance to umbilical point 1 ($d_{VDU,U1}$) (a, c; normalized by geodesic distance between umbilics) and the corresponding vortex order parameter (b, d; $\Phi$) on prolate (a, b; $x/z=0.25$) and oblate spheroids (b, d; $x/z=4$) for 4 different particle numbers and $v_{0}=0.5$. The time required to reach stability increases with the particle number and takes significantly longer on oblate spheroids compared to prolate spheroids.}
\end{figure*}

%\section{Discussion}
\textit{Discussion} - This work identified umbilical points on ellipsoidal surfaces as crucial geometric features to interpret collective motion patterns on closed surfaces. Umbilical points define special surface regions of high geodesic separation and provide information about local variation in curvature. We have shown that vortex motion is connected to these umbilics, where normal curvature is constant. To explain the observed motion patterns, we need to consider interactions between defects (e.g., vortices) in the director field, interactions between these defects and geometric features of the surface, as well as dynamic effects from bulk particle motion. It is known that vortices repel each other with an interaction energy depending linearly on separation distance \cite{bowick2009two, reuther2015interplay}. On surfaces with non-constant Gaussian curvature, each vortex experiences an additional geometric potential determined by the local Gaussian curvature \cite{turner2010vortices}. In this case, the vortex interactions can be described by an effective free energy \cite{vitelli2004anomalous}, that takes into account the broken translational invariance due to intrinsic curvature. This energy essentially describes the deviation from perfect alignment in the vector-field and implies that the energy of the system is minimised when the particle alignment is globally maximised. Although the energy of the vortex in our system is not clearly defined, these concepts can still help us to understand the vortex dynamics around umbilics in the simple cases of prolate and oblate spheroids (Fig. 3a).

%\onecolumngrid
\begin{figure*}
\includegraphics[width=0.8\linewidth]{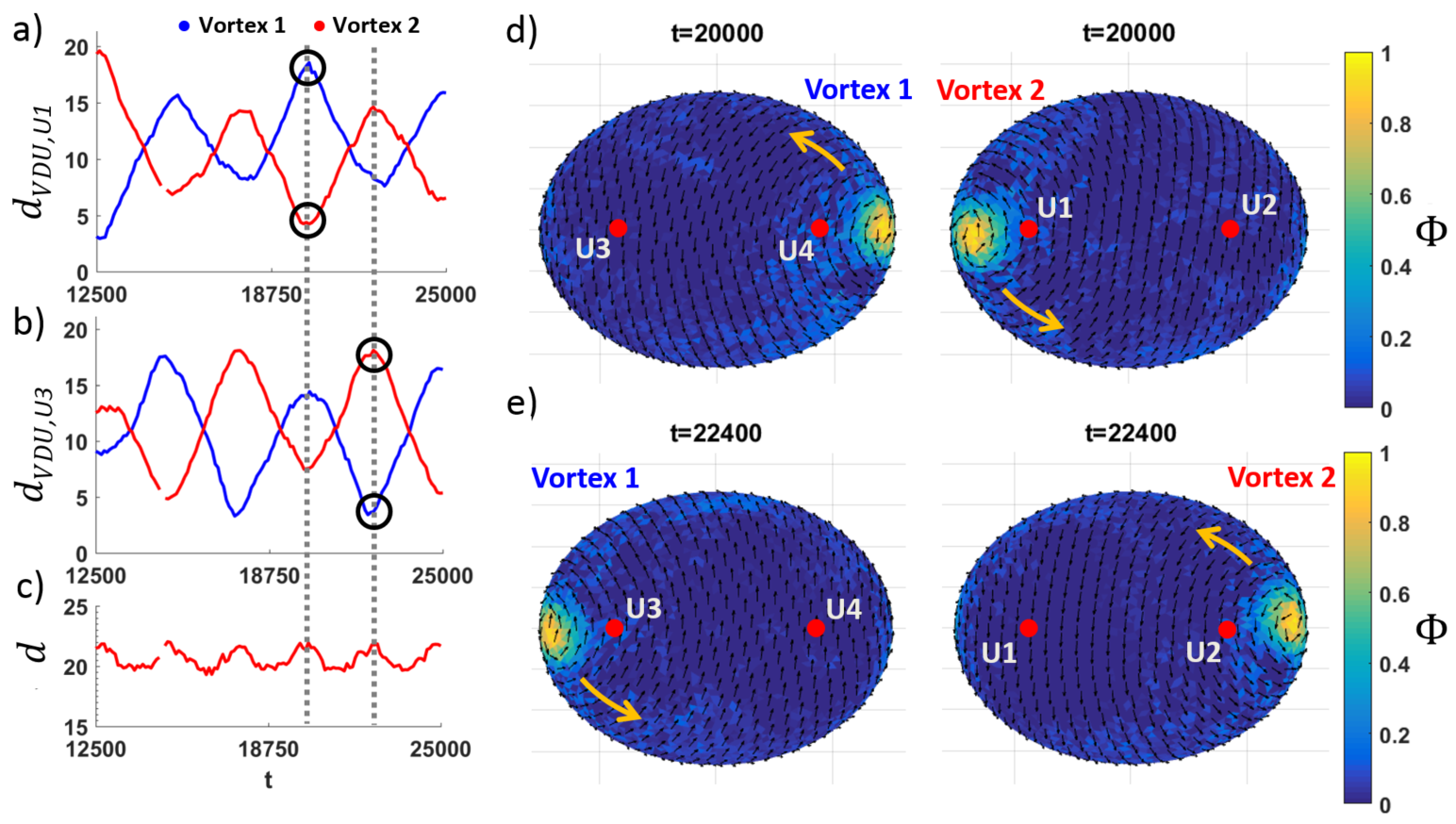}
\caption{\label{VDU_examples_triaxial_ellipsoids} Vortex dynamics on a triaxial ellipsoid with axis ratios $x/z=2$, $x/y=1.14$ and $v_{0}=0.1$. a) and b) show the $d_{VDU}$ for vortices 1 and 2 (blue and red curves) as a function of time measured from two different umbilical points (U1, U3). The circles indicate the maximum distance of the vortices at $t=2({10}^{4})$ (a) and $t=2.24({10}^{4})$ time-steps (b), which are peak values of the vortex to vortex distance shown in c). Images depicted in d) and e) are the corresponding mapped values of $\Phi$ and director-fields. On triaxial-ellipsoids with pairs of close umbilical points the vortices encircle the two closest umbilics whereas they switch positions when umbilics are further apart. All cases lead to stable oscillations in distance as illustrated in (a) and (b). The distance between vortices is maximised close to the umbilical points and switches between the two symmetric configurations (c).}
\end{figure*}
%\twocolumngrid

On prolate spheroids the location of umbilics coincide with regions of high Gaussian curvature (and geometric potential), causing vortices to be pushed towards umbilics, since it increases the global alignment of the director field. This approach of the vortex towards a point of high Gaussian curvature at the same time reduces the local alignment of the vortex vector field, causing an avoidance of the umbilics. Vortex dynamics thus arise from a balance between these opposing factors. With increasing aspect ratio ($x/z \ll 1$) the contribution of the global alignment becomes predominant leading to decreasing $d_{VDU}$ values (Fig. 3a). Using the same reasoning, on oblate spheroids we would expect that the high Gaussian curvature rim will be avoided by vortices, while at the same time the higher global alignment that can be achieved in the flatter region will be obtained when the vortex approaches the rim. The alignment of particles moving parallel to the rim, however, is increased when the vortex is located at the umbilical point. This alignment becomes further increased at higher particle velocities, leading to decreasing $d_{VDU}$ values (Fig. 3a). The dependence of $d_{VDU}$ on aspect ratio, and a quantitative understanding of the orbital frequency of vortex motion around the umbilic, however cannot be explained using this energetic argument. Additional insight can be gained by a simple approximation of spheroids as capped cylinders \cite{turner2010vortices}, where the vortex interaction energy, E is proportional to H/R, where R is the radius and H is the height of the capped cylinder. This simple approximation immediately implies that the interaction energy is lower on oblate spheroids compared to prolate spheroids and explains why the $d_{VDU}$ in Fig. 2 is larger on oblate spheroids. The geometric potential of the umbilics decreases with decreasing aspect ratio and hence the vortices are less constrained, which is reflected by the increasing $d_{VDU}$ in Fig 3a.

%\onecolumngrid
\begin{figure}[h]
\includegraphics[width=\linewidth]{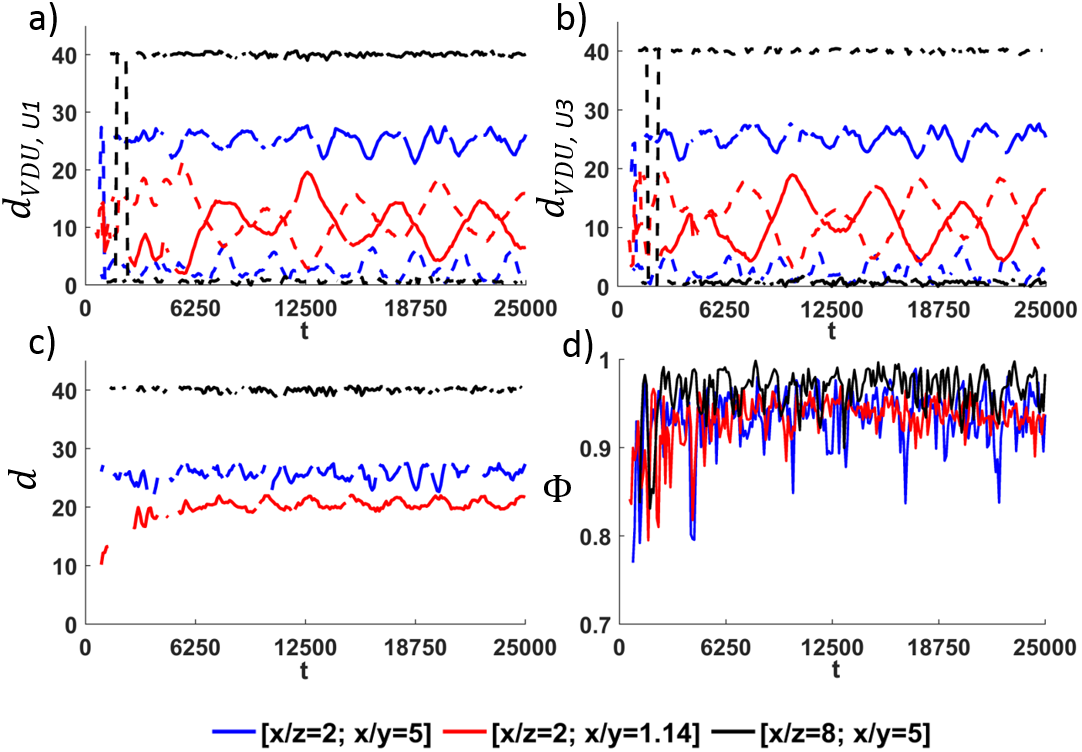}
\caption{\label{VDU_evolution_triaxial_ellipsoids} Evolution of the vortex distance to umbilical point ($d_{VDU}$) for three different triaxial ellipsoids with axis ratios $x/z=2$, $x/y=5$ (blue); $x/z=2$, $x/y=1.14$ (red) and $x/z=8$, $x/y=5$ (black) and $v_{0}=0.1$. a) and b) show the distance to the umbilical points ($d_{VDU}$, U1, U3) for each of the two vortices (dashed line indicates second vortex). The distance between the vortices is maximized with fewer fluctuations on elongated-ellipsoids (c). Vortices quickly form and are stable for the entire simulation time with $\Phi$ values well above the threshold value of 0.7 (d)}
\end{figure}
%\twocolumngrid

The further loss of symmetry on triaxial ellipsoids adds some additional complexity to the interactions between vortices and surface geometry. The two pairs of symmetric umbilical points still define a low energy configuration of the system since they define the positions of maximum separation distance for the vortices (Fig. 6c). Thus the $d_{VDU}$ shows only small fluctuations on prolate-like and oblate-like ellipsoids with pairs of close umbilics (Fig. 7a, b). The energy in the vector-field decreases with increasing velocity due to increasing alignment and causes vortices to be further attracted to high Gaussian curvature regions between the umbilics (suppl. mat. movie 5 \cite{Supplemental_Movies}). In the case of flat ($x/y>3$ and $x/z<1$) triaxial ellipsoids no vortices were observed (triangles in Fig. 3b). This is because the potential energy stored in the vector-field can only partly be minimised by rotational motion, which leads to motion patterns that quickly change orientations at the poles. In contrast, on oblate spheroids of high aspect ratio, particles are still able to form vortices since they align with the sharp edge of the ellipsoid.\\
The dynamical features that we presented in this manuscript have significant implications on understanding the behaviour of numerous active biological systems where curvature plays an important role. In the cell membrane, for instance, complex cellular processes such as cell signaling and shape regulation of organelles rely on the collective dynamics of molecules confined within a two dimensional curved lipid-bilayer \cite{alberts2007molecular}. The curvature of the membrane thereby effects the interaction and dynamics of the molecules leading to dynamic patterns that greatly differ from their flat euclidean counterpart \cite{sigurdsson2016hydrodynamic}. On the tissue level, cell migration occurs on curved tissues of the intestinal crypt \cite{fatehullah2013cell, ritsma2014intestinal} and has been observed during morphogenesis of the mammary epithelia where the collective rearrangement of cells determine architecture and polarity of the epithelia. Large flows of collectively migrating cells constrained to move on a curved surface occur during gastrulation of the chick embryo \cite{vasiev2010modeling} and during embryonic development of the zebra fish embryo \cite{keller2008reconstruction}. Curvature also effects the collective migration of cells during the development of the corneal epithelia leading to vortex pattern of radial stripes \cite{collinson2002clonal}.\\
Despite intensive research, the mechanisms by which these cell movements are orchestrated to form structures much larger than the individual cell remain poorly understood. In tissues, collective motion emerges as a result of direct physical contact as has been shown for cells moving on flat substrates \cite{szabo2006phase, trepat2009physical}. These experiments, however, neglect the fact that in reality cells are constrained to move on curved surfaces. Our simple model of polar active particles confined to move on ellipsoidal surfaces provides new insights into how curvature affects motion patterns in active systems. The aim was to explore the effect of varying curvature and topology on collective motion. We have thereby shown that active directed motion and intrinsic surface curvature lead to complex motion patterns: particles tend to move along geodesic paths and are strongly influenced by topological constraints resulting in particles encircling surface points of constant curvature. These observations may help to understand the underlying mechanisms of self-organization and collective cell migration on non-constant Gaussian curvature surfaces such as the coordinated cell migration during the development of the zebra fish embryo \cite{keller2008reconstruction}.

\textit{Conclusion} - In summary, we have explored how geometry affects the collective behaviour of active particles confined to move on a curved surface. The non-linear coupling between non-constant Gaussian curvature and defect-defect interactions gives rise to a variety of motion patterns that can be partially interpreted by theories of vortex-geometry interactions. The richness of physics observed in our study can be expected to further increase if one of the following constraints is released: (i) a reduction of the packing fraction leaving “more space” for the particles, (ii) a softer interaction between the particles allowing large particle overlaps and (iii) surfaces with gradients of positive and negative Gaussian curvature that have isolated or odd numbers of umbilical points, i.e. handles. Our results suggest that Gaussian curvature may also be responsible for the emergence of complex patterns in a variety of active systems, such as collective cell behaviour during morphogenesis.\\

\begin{acknowledgments}
We thank Peter Fratzl for stimulating discussion, and acknowledge funding for SE, from the Leibniz prize of Peter Fratzl running under DFG contract number FR2190/4-1.  JF, RW, and JD were supported by the German Research Foundation in the Cluster of Excellence Interdisciplinary Laboratory “Image Knowledge Gestaltung” (DFG Contract No. 415 EXC1027/1).
\end{acknowledgments}

\bibliography{Ehrig_et_al_2016}{}

\bibliographystyle{ieeetr}
   
\end{document}